\documentclass[sigconf]{acmart}
\AtBeginDocument{%
  }

\usepackage{xspace}
\usepackage{subcaption}
\usepackage{caption}
\usepackage{graphicx} 
\usepackage{booktabs} 
\usepackage{multirow}

\copyrightyear{2026}\acmYear{2026}
\setcopyright{none}
\acmConference[AISec '26]{19th Workshop on Artificial Intelligence and Security}{November 15-19, 2026}{The Hague, Netherlands}
\acmBooktitle{19th Workshop on Artificial Intelligence and Security (AISec '26), November 15-19, 2026, The Hague, Netherlands}
\acmDOI{10.1145/3847352.3848088}
\acmISBN{979-8-4007-3031-3/2026/11}

\newcommand{\sys}{\textsc{HideOut}\xspace}

\newcommand\blfootnote[1]{%
  \begingroup
  \renewcommand\thefootnote{}\footnote{#1}%
  \addtocounter{footnote}{-1}%
  \endgroup
}

\newcommand{\aco}[1]{\textcolor{red}{#1}}
\newcommand{\mic}[1]{\textcolor{cyan}{#1}}

\newcommand{\stepnum}[1]{%
  \tikz[baseline=(step.base)]%
    \node[circle,fill=black,text=white,inner sep=0pt,minimum size=0.9em,
          font=\scriptsize\bfseries] (step) {#1};%
}
\begin{document}
\makeatletter
\newcommand{\setsublabel}[2]{%
  \protected@edef\@currentlabel{\the\numexpr\value{figure}+1\relax(#2)}%
  \label{#1}%
}
\makeatother
\title{\sys{} - User Protection Against Localization with Throughput Guarantees in Mobile Networks}

\author{Navidreza Asadi}
\authornote{Equal contribution.}
\email{navidreza.asadi@tum.de}
\orcid{0000-0002-2155-1288}
\affiliation{%
  \institution{Technical University of Munich}
  \city{Munich}
  \country{Germany}
}

\author{Michele Guardiani}
\email{michele.guardiani@tum.de}
\orcid{0009-0003-9116-6929}
\authornotemark[1]
\affiliation{%
  \institution{Technical University of Munich}
  \city{Munich}
  \country{Germany}
}

\author{Wolfgang Kellerer}
\email{wolfgang.kellerer@tum.de}
\orcid{0000-0003-4358-8038}
\affiliation{%
  \institution{Technical University of Munich}
  \city{Munich}
  \country{Germany}}

\author{Jafar Aco Mohammadi}
\email{jafar.mohammadi@nokia.com}
\orcid{0009-0000-7141-2107}
\affiliation{%
  \institution{Nokia Research \& Innovations}
  \city{Munich}
  \country{Germany}}

\renewcommand{\shortauthors}{Asadi et al.}

\begin{abstract}
Channel state information (CSI) in cellular networks is primarily used to maintain quality of service (QoS), yet it leaks privacy-sensitive information, in particular the user equipment's (UE) location. 
CSI is reported in plaintext on a publicly available schedule; thus, capturing it on the air and using a well trained AI model to localize the UE is straightforward. This simplicity in the attack execution makes it a serious privacy vulnerability. Using conventional privacy enhancing techniques, such as additive perturbation from differential privacy,  to protect location data can potentially degrade the downlink communication quality.

We present \sys{}, a client-side defense mechanism that preserves the user's location from neural-based localization methods, while maintaining their downlink Spectral Efficiency (SE).
Unlike generic adversarial attacks, \sys treats downlink SE as a strict constraint in an optimization setting. 
Under maximum ratio transmission (MRT) beamforming, we show that a user-selected SE loss budget can be converted into an angular rate-feasible constraint on the reported channel vector. \sys then performs gradient-based localization obfuscation while projecting each update onto this rate-feasible set.
We evaluate \sys against four CSI-based localization architectures under both \textit{white-box} and \textit{black-box} assumptions. 
Across models, \sys increases localization ambiguity (error) from approximately $1$--$2$\,m to $72$--$85$\,m while keeping downlink SE loss within a $1\%$ budget. These results show that CSI-based localization can be substantially degraded at negligible communication quality loss when adversarial perturbations are made aware of the wireless link.

\end{abstract}






\maketitle

\blfootnote{\copyright{} 2026 Navidreza Asadi, Michele Guardiani,
Wolfgang Kellerer, and Jafar Aco Mohammadi. This is the author's
version of the work. It is posted here for your personal use. Not for
redistribution. The definitive version was published in
\emph{Proceedings of the 19th Workshop on Artificial Intelligence and
Security (AISec~'26), November 15--19, 2026, The Hague, Netherlands},
\url{https://doi.org/10.1145/3847352.3848088}}

\section{Introduction}

\begin{figure}[t!]
    \centering
    \includegraphics[width=0.9\columnwidth]{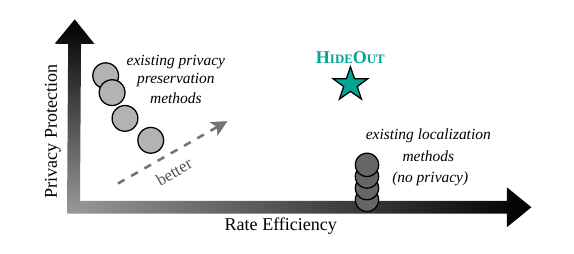} 
    \caption{Localization Privacy-Protection vs. Service Quality (Rate Efficiency) Trade-off.}
    \label{fig:hideout_advantage}
    \Description{Illustration of the tradeoff between localization privacy-protection and service quality (rate efficiency). The x-axis represents the rate efficiency, while the y-axis represents the level of localization privacy-protection. The graph shows that as localization privacy-protection increases, there is a potential decrease in rate efficiency, highlighting the tradeoff between the two objectives. We propose \sys that not only preserves localization privacy but also maintains a high level of rate efficiency, thus achieving an optimal balance between the two.}
\end{figure}

Cellular networks are becoming sensing systems as well as communication systems. 
In 5G-Advanced and emerging 6G deployments, the same radio measurements that support beamforming and link adaptation can also be used to infer a User Equipment's (UE's) location. 
The 3GPP architecture already includes a Location Management Function (LMF) that collects positioning-related measurements from UEs~\cite{3gpp_tr22261_rel19_2024}. 
Recent learning-based positioning systems further strengthen this capability. Neural networks can map channel state information (CSI) directly to UE coordinates, exploiting multipath, blockage, and other propagation effects that are highly location-dependent~\cite{colpaert2023massive,vuckovic2024csi,pan2025ai}. 
Thus, CSI is no longer only a communication primitive, but also a high-resolution location signal.

This dual use creates a privacy risk. A
 UE must report CSI to receive efficient downlink service, but that same report may allow a network-side localization service to infer the UE's precise position. 
Such location information can reveal sensitive attributes such as home and workplace, mobility routines, social relationships, and visits to sensitive places~\cite{wang2022srr,zhou2025feasibility}. 
Protecting only the location record after LMF is therefore insufficient: if the localization service can infer location directly from CSI, then privacy must be enforced before or during CSI reporting.

Existing location privacy mechanisms do not address this leakage channel. 
Coordinate-level defenses such as $k$-anonymity, geo-indistinguishability, and local differential privacy perturb or generalize a location after it has already been estimated~\cite{andres2013geo,wang2022srr}. 
Signal-level defenses are closer to our setting, but most prior work targets WiFi or other short-range sensing systems, where preserving communication often means only that packets remain decodable~\cite{cominelli2020csi,zhou2020adversarialwifi,xiao2023fooloc,li2024practicalwifi}. 
However, this assumption does not carry over to cellular downlink transmission, as the base station (BS) uses the reported CSI to compute the precoder, making the achievable rate depend continuously on CSI fidelity.

This leads to the central challenge of this paper (Fig.~\ref{fig:hideout_advantage}): a UE cannot simply randomize or suppress its CSI to hide its location. 
Perturbing CSI may mislead a learning-based localization model, but it can also misalign the downlink beamformer and reduce the UE's own rate. 
The privacy mechanism must therefore satisfy two requirements simultaneously: it should increase the localization error of the network-side model, while guaranteeing that the protected CSI remains useful for downlink precoding. 
This privacy--service coupling is specific to cellular CSI feedback and is not captured by generic adversarial attacks or by prior WiFi-oriented signal-level privacy mechanisms.

\begin{figure}[!t]
    \centering
    \includegraphics[width=0.9\columnwidth]{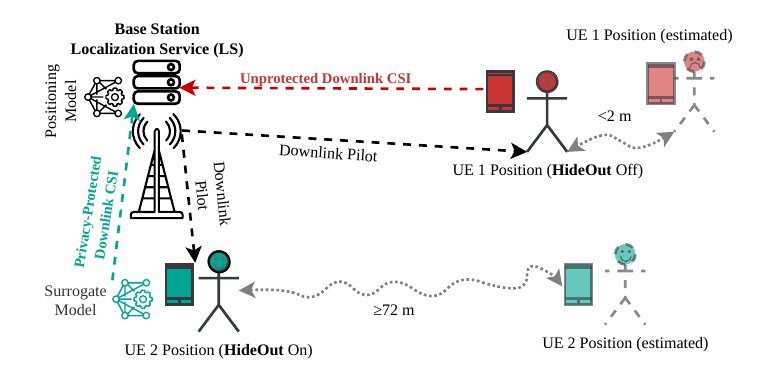} 
    \caption{Overall workflow. UE2 protects its location using \sys{} by inducing an error greater than $72$\,m while UE1's location is estimated at granularity of less than $2$\,m.
    }
    \label{fig:workflow}
    \Description{System architecture diagram comparing unprotected and privacy-protected UE localization frameworks.}
\end{figure}

We propose \sys, a communication-aware CSI perturbation mechanism for protecting UE location against ML-based cellular localization. 
As depicted in Fig.~\ref{fig:workflow}, the UE reports a perturbed CSI matrix instead of its original measurement. 
The perturbation is optimized to increase the localization loss of a neural positioning model, but each update is projected onto a wireless-feasible set that guarantees a user-selected bound on downlink SE loss. 
Mathematically, we show that under Maximum Ratio Transmission (MRT) precoding, a bound on SE degradation can be converted into an angular constraint on the reported channel direction. 
This yields a projection-based mechanism whose rate feasibility is enforced by construction, rather than by tuning a soft penalty term.

\sys supports both \textit{white-box} and \textit{black-box} localization services. 
In the \textit{white-box} setting, the UE computes perturbations using gradients of the deployed localization model. 
In the more realistic \textit{black-box} setting, the localization model is proprietary. The UE trains a local surrogate from its own CSI-location pairs and relies on transferability of adversarial perturbations~\cite{papernot2017practical,liu2017delving}. 
Unlike standard \textit{black-box} adversarial attacks, however, \sys must satisfy a physical layer constraint in which every protected CSI report must remain sufficiently aligned for downlink precoding.

In summary, our paper's contributions are: 





    \noindent \textbf{$\blacktriangleright$~CSI feedback as a privacy attack surface.}
    We formulate location privacy in mobile networks as a client-side defense problem over CSI feedback. The formulation captures a distinctive privacy--service coupling. The same CSI report is used by the localization service to infer position and by the base station to compute the downlink precoder.

    \noindent \textbf{$\blacktriangleright$~Rate-constrained adversarial CSI perturbations.}
    We derive an angular feasibility condition that converts a SE budget into a constraint on the reported channel direction under MRT precoding. 
    \sys uses this condition to project gradient-based perturbations onto a rate-feasible set.

    \noindent \textbf{$\blacktriangleright$~\textit{Black-Box} protection via surrogate transfer.}
    We design a \textit{black-box} version of \sys in which the UE trains a surrogate localizer through Bayesian architecture search, without access to the deployed model's architecture, parameters, predictions, or gradients.

    \noindent \textbf{$\blacktriangleright$~Evaluation across localization architectures.}
    We evaluate \sys across four positioning models and both \textit{white-box} and \textit{black-box} settings. 
    In our experiments, \sys increases localization error from approximately $1$--$2$\,m to $72$--$85$\,m while keeping the downlink rate loss within $1\%$ of the unprotected baseline.

\section{Background}
\textbf{Notation.} Matrices and vectors are denoted by boldface upper-case and lower-case letters, respectively. All vectors are defined as column vectors. The notations $(\cdot)^T$, $(\cdot)^H$, and $\|\cdot\|$ denote transpose, conjugate transpose, and the $\ell_2$ norm, respectively. The inner product between two complex vectors is defined as $\langle \mathbf{a}, \mathbf{b} \rangle = \mathbf{b}^H \mathbf{a}$.

\subsection{A Primer on the Physical Layer}
We consider a Frequency Division Duplexing (FDD) Single-User (SU) Multiple-Input Single-Output (MISO) downlink system. Under FDD, the uplink and the downlink transmissions are on separate frequency bands. In MISO configurations, the BS is equipped with $N_{\text{Tx}}>1$ elements, while the UE employs a single isotropic antenna. The received signal $y_k \in \mathbb{C}$ at the UE is defined as
\begin{equation}
y_k = \mathbf{h}^H_k \mathbf{w}_k s_k + n_k,
\end{equation}
\noindent where $\mathbf{h}_k \in \mathbb{C}^{N_{\text{Tx}}}$ is the channel vector for the $k$-th subcarrier, $\mathbf{w}_k \in \mathbb{C}^{N_{\text{Tx}}}$ and $ s_k \in \mathbb{C}$ are the transmit precoder vector and the transmit symbol and $n_k \sim \mathcal{CN}(0, \sigma^2)$ represents additive white Gaussian noise. 
Given the FDD operation mode, the base station cannot infer the downlink channel and must rely on the UE feedback. Thus, the base station transmits Channel State Information Reference Signal (CSI-RS) pilot symbols ~\cite{3gpp38211}. 
The UE estimates the true downlink channel as 
\begin{equation}
    \mathbf{\hat{H}} = [\mathbf{\hat{h}}_1, \dots, \mathbf{\hat{h}}_K] \in \mathbb{C}^{N_{\text{Tx}} \times K}.
\end{equation}
We then assume that the UE reports the estimated channel $\mathbf{\hat{H}}$ as CSI feedback without compression.
Given $\mathbf{\hat{H}}$, the base station computes the precoding vector and transmits the signal $\mathbf{x}_k = \mathbf{w}_k s_k$, where $s_k\in \mathbb{C}$ is the data symbol. The beamforming vector $\mathbf{w}_k$ is thus a function of the reported CSI, i.e. $\mathbf{w}_k := \mathbf{w}\left(\mathbf{\hat{h}}_k\right)$. More specifically, for the considered scenario, the optimal beamformer vector is the MRT beamformer \cite{mrtBeam}, defined as 
\begin{equation}
    \label{eq:mrt}
    \mathbf{w}\left(\mathbf{\hat{h}}_k\right) :=\sqrt{p} \frac{\mathbf{\hat{h}}_k}{\left\|\mathbf{\hat{h}}_k\right\|}, 
\end{equation}
%
%
where $p$ is the transmit power, which we assume to be equal to one for simplicity.
Thus, for a given bandwidth and SNR level, the maximum achievable SE is a function of the precoding vector and is defined as
\begin{equation}
    \label{eq:spectralEff}
    \eta(\mathbf{\hat{h}_k}) := \log_2 \left( 1 + \frac{|\langle \mathbf{h}_k \, , \,  \mathbf{w}_k  \rangle|^2}{\sigma^2} \right),
\end{equation} 
where $\sigma^2$ is the noise power. For a given bandwidth and SNR level, $\eta\left(\cdot\right)$ sets a fundamental limit on the per-subcarrier maximum rate.

\subsection{ML-based Positioning}
\label{sec:ml-positioning}


We assume that at the LS a neural network (NN)-based positioning model $f_\theta(\cdot)$ is deployed which,
given the CSI reported by the UE, estimates the UE's two dimensional coordinates as
\begin{equation}
    \hat{\mathbf{r}} = f_{\theta}(\mathbf{\hat{H}}).
\end{equation}
The ground-truth is denoted by $\mathbf{r} = [r_x,r_y]^T \in \mathbb{R}^{2}$. We train the model $f_{\theta}(\cdot)$ as a regression function using pairs of CSI measurements and UE coordinates. The training loss function is the mean squared error between $\hat{\mathbf{r}}$ and $\mathbf{r}$, while we report positioning accuracy as the Euclidean error $\|\hat{\mathbf{r}}-\mathbf{r}\|$.

The model receives the real and imaginary components of $\mathbf{\hat{H}}$ as two input channels. It then derives four additional features: magnitude, phase, and the real and imaginary components of the inverse discrete Fourier transform along the subcarrier dimension. This produces a six-channel representation that exposes information from both the frequency and delay domains to the positioning model. Section~\ref{sec:localization-models} describes the model architectures used in our evaluation.






\begin{figure*}[t!]
    \includegraphics[width=1.0\textwidth]{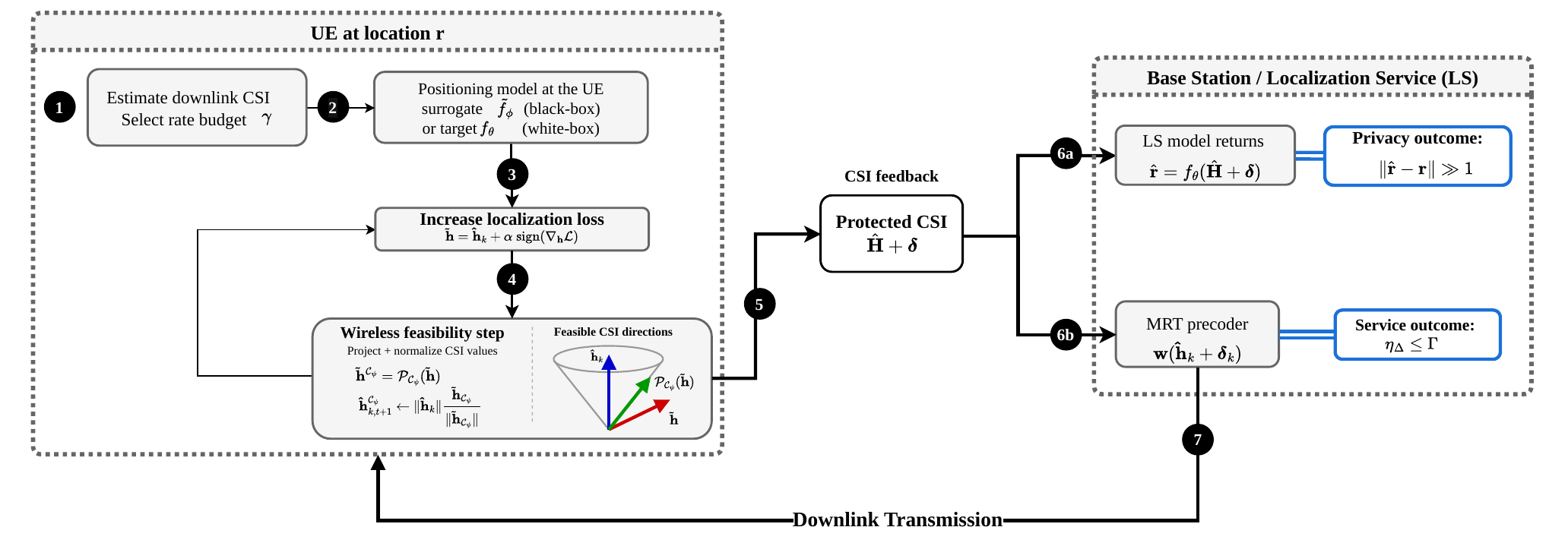}
    \caption{\sys{} for a CSI reporting instance.
    The inset in Step~4 illustrates the change from the estimated to the protected CSI and the projection of an unconstrained update onto the cone of communication-feasible directions. The resulting report causes a large localization error in the localization service at the base station while retaining a bounded SE loss for downlink communication.}
    \Description{Illustration of HideOut integrated within a CSI reporting cycle. A UE at a new location, estimates the channel, computes the perturbation and reports to the BS the perturbed CSI. The BS, given the report, infers the wrong UE location but correctly beamforms the downlink signal. }
    \label{fig:hideout-scenario}
\end{figure*}

\section{Problem Formulation}
\label{sec:threat-model}

This section states the trust assumptions, the protected asset, and the knowledge available to each party in the CSI reporting loop introduced in Section~\ref{sec:ml-positioning}. Section~\ref{sec:AA4LP} uses these assumptions to derive the privacy mechanism itself.

\noindent\textbf{$\blacktriangleright$} \textit{Protected asset.} 
The protected asset is the UE's precise two-dimensional location $\mathbf{r}$. We do not attempt to hide the UE's presence in the cell, its association with the serving base station, or coarse network-level reachability. Instead, we aim to prevent the localization service from recovering fine-grained coordinates from the CSI report that the UE must provide for downlink communication.

\noindent\textbf{$\blacktriangleright$} \textit{Trust model.} The UE does not trust the base station or its LS with its location: any CSI report may be used by $f_\theta$ to estimate its coordinates, regardless of the UE's privacy preferences. The UE only relies on the base station to follow the standard communication protocol, i.e., to decode the reported CSI and compute a precoder from it, since this is required to receive service at all.

\noindent\textbf{$\blacktriangleright$} \textit{Defender capabilities.} 
The UE controls the CSI values it reports and knows its own channel estimate and location during defense construction. It cannot change the physical propagation channel or the localization model deployed by the network. 
We evaluate the privacy mechanism under two knowledge settings. In the \textit{white-box} setting, the UE has access to the target model $f_{\theta}$ deployed at the LS and computes perturbations using its gradients.
In the more realistic \textit{black-box} setting, the base station LS keeps its model secure. The UE trains a local surrogate $\tilde{f}_{\theta}$ using CSI-location pairs. It computes the perturbation from the surrogate and relies on transferability to the unknown target model. Section~\ref{sec:surrogate-search} describes how we construct the surrogate models without using the target model's architecture, parameters, predictions, or gradients.

\noindent\textbf{$\blacktriangleright$} \textit{LS knowledge and constraints.} The LS applies the positioning function $f_\theta$ (Section~\ref{sec:ml-positioning}) to any CSI it receives.
It does not run 
adversarial training, or fuse the CSI with an independent, trusted positioning source to recover the UE's location, and it does not deny service or otherwise change its precoding procedure in response to localization uncertainty. 
LS's architecture and parameters $\theta$ are proprietary and unavailable to the UE by default; in the \textit{white-box} setting, however, we relax this assumption to evaluate the benefit of full model knowledge.
\section{\sys{} Methodology}
\label{sec:AA4LP}

\subsection{Location Obfuscation by Adversarial Perturbation}

The UE protects its location by constructing a perturbation $\boldsymbol{\delta} \in \mathbb{C}^{N_{\text{Tx}} \times K}$ and reporting $\mathbf{\hat{H}}+\boldsymbol{\delta}$ instead of the original channel estimate. An optimal perturbation is found by jointly encoding the privacy and service objectives. The localization objective maximizes the regression loss of the positioning model, while the SE constraint limits the degradation caused by mismatched beamforming. 
The resulting optimization does not seek the largest unconstrained positioning error. It seeks the largest error that the UE can obtain within an explicit communication budget.

\noindent\textbf{Optimization Framework.}
The optimal perturbation $\boldsymbol{\delta}$ is thus obtained by solving 

\begin{equation}
\label{eq:optimization_problem}
\begin{aligned}
\boldsymbol{\delta} = \arg\max_{\boldsymbol{\delta}} \quad & \left\| f_{\theta}\left(\mathbf{\hat{H}} + \boldsymbol{\delta}\right) - \mathbf{r} \right\|^2 \\
\text{s.t.} \quad & \eta_{\Delta} \leq \Gamma_k, \\
&\left\|\mathbf{\hat{h}}_k + \boldsymbol{\delta}_k\right\| = \left\|\mathbf{\hat{h}}_k\right\|, \quad \forall k,
\end{aligned}
\end{equation}
\noindent where $\eta_\Delta$ measures the impact of mismatched beamforming on the SE and is defined as
\noindent 
\begin{equation}
    \label{eq:effDelta}
    \eta_{\Delta} := \eta\left(\mathbf{\hat{h}_k}\right) - \eta\left(\mathbf{\hat{h}_k} +\boldsymbol{\delta}_k\right).
\end{equation}
The parameter $\Gamma_k$ denotes the maximum permissible degradation, and the normalization constraint ensures that the perturbed channel preserves the original power profiles. In the rest of the paper we express the degradation budget as a fraction of the rate the UE would obtain without protection, i.e., $\Gamma_k = \gamma\,\eta(\hat{\mathbf{h}}_k)$, and report the user-selected $\gamma \in (0,1)$ as a percentage.

\noindent\textbf{Geometrical Interpretation.}
Defining the Hermitian angle between two complex vectors as the arccosine of the absolute value of their normalized inner product, we can show that the constrained perturbation problem in Eq. \eqref{eq:optimization_problem} is equivalent to 
\begin{equation}
\label{eq:constrainedDirection}
\begin{aligned}
\boldsymbol{\delta} = \arg\max_{\boldsymbol{\delta}} \quad &  \left\|  f_{\theta}\left(\mathbf{\hat{H}} + \boldsymbol{\delta}\right) - \mathbf{r} \right\|^2 \\
\text{s.t.} \quad & |\langle \mathbf{h}_k, \mathbf{w}(\mathbf{\hat{h}}_k + \boldsymbol{\delta}_k) \rangle|\geq \|\mathbf{h}_k\|\cos \psi_k, \\
& \left\|\mathbf{\hat{h}}_k + \boldsymbol{\delta}_k\right\| = \left\|\mathbf{\hat{h}}_k\right\|, \quad \forall k,
\end{aligned}
\end{equation}
where the angular threshold $\psi_k$ is defined as
\begin{equation}
    \psi_k = \arccos \left( \sqrt{\frac{\left(1+\rho_k \cos^2\beta_k\right)^{1-\gamma} -1}{\rho_k}} \right),
\end{equation}
$\beta_k$ is  the Hermitian angle between the true channel $\mathbf{h}_k$ and its estimate $\mathbf{\hat{h}}_k$ and $\rho_k := \|\mathbf{h}_k\|^2/\sigma^2$ denotes the SNR of the true channel when the precoder matches the channel. The equivalence follows from the monotonicity of the SE $\eta$, which is a strictly increasing function of the received signal power. A more detailed derivation can be found in the Appendix~\ref{apdx:gammaPsi}.

\subsection{Wireless Context Awareness}

Eq. \eqref{eq:constrainedDirection} restricts the optimal perturbation to be within the feasible set 
\begin{equation}
    \mathcal{C}_\psi(\mathbf{\hat{h}}_k) = \left\{\mathbf{x} \in \mathbb{C}^{N_{\text{Tx}}} \;\middle|\; |\langle \mathbf{x}, \mathbf{\hat{h}}_k \rangle | \ge \|\mathbf{x}\|\,\|\mathbf{\hat{h}}_k\| \cos\psi_k \right\},
\end{equation}
\noindent i.e., a complex cone $\mathcal{C}_\psi(\mathbf{\hat{h}}_k)$ with axis $\mathbf{\hat{h}}_k$ and aperture $\psi_k$. We thus propose \sys, a gradient-based approach specifically designed for the considered wireless scenario which finds a perturbation that maximizes a localization model loss while ensuring that the reported CSI remains within $\mathcal{C}_\psi(\mathbf{\hat{h}}_k)$ therefore satisfying the downlink SE constraints.

\noindent\textbf{The \sys Optimization Algorithm.}
\sys, outlined in Algorithm~\ref{alg:hideOutAlgo}, consists of $T$ iterations of three key sub-steps: an initial gradient ascent step, a projection step of the perturbed sample onto $\mathcal{C}_\psi(\mathbf{\hat{h}}_k)$ and lastly a normalization step.

\begin{algorithm}
\caption{\sys}
\label{alg:hideOutAlgo}
\renewcommand{\baselinestretch}{1.5}
\renewcommand{\algorithmicrequire}{\textbf{Input:}}
\renewcommand{\algorithmicensure}{\textbf{Output:}}
\renewcommand{\baselinestretch}{1.5}
\begin{algorithmic}[1]
\REQUIRE $\mathbf{\hat{h}}_k,\;\mathbf{r},\;T,\;\psi_k,\;\alpha$ 
\ENSURE $\mathbf{\hat{h}}_{k,T}^{\mathcal{C}_\psi} \in \mathcal{C}_\psi(\mathbf{\hat{h}}_k)$ with $\left\|\mathbf{\hat{h}}_{k,T}^{\mathcal{C}_\psi}\right\| = \left\|\mathbf{\hat{h}}_{k}\right\| $
\STATE $\mathbf{\hat{h}}_{k,0}^{\mathcal{C}_\psi} \gets \mathbf{\hat{h}}_k$

\FOR{$t = 0$ to $T-1$}
    \STATE $\mathbf{g}_t \gets \nabla_{\mathbf{h}} \mathcal{L}(\mathbf{\hat{h}}_{k,t}^{\mathcal{C}_\psi}, \mathbf{r}; \theta)$ 
    \STATE $\tilde{\mathbf{h}} \gets \mathbf{\hat{h}}_{k,t}^{\mathcal{C}_\psi} + \alpha\;\mathrm{sign}(\mathbf{g}_t)$\label{line:gradient_ascent}
    \STATE $\tilde{\mathbf{h}}^{\mathcal{C}_\psi} \gets \mathcal{P}_{\mathcal{C}_\psi}(\tilde{\mathbf{h}})$    \label{line:projection}
    \STATE $\mathbf{\hat{h}}_{k,t+1}^{\mathcal{C}_\psi} \gets \|\mathbf{\hat{h}}_k\|\frac{\tilde{\mathbf{h}}_{\mathcal{C}_\psi}}{\|\tilde{\mathbf{h}}_{\mathcal{C}_\psi}\|}$\label{line:normalization}
    
\ENDFOR

\STATE \textbf{return} $\mathbf{\hat{h}}_{k,T}^{\mathcal{C}_\psi}$
\end{algorithmic}
\end{algorithm}


The initial gradient step, in line ~\ref{line:gradient_ascent} of Algorithm~\ref{alg:hideOutAlgo}, takes a step of size $\alpha$ from the adversarial sample in the direction that maximizes the localization loss $\mathcal{L}(\cdot)$. The projection $\mathcal{P}_{\mathcal{C}_\psi}(\cdot)$ onto the cone in line ~\ref{line:projection} guarantees that the perturbed sample remains within the feasible set, thus satisfying the SE constraint. The projection consists of first applying a phase rotation to render the Hermitian inner product real-valued, performing a real second-order cone projection, and then rotating the result back to the original complex phase. A mathematical description of the projection operation can be found in the Appendix~\ref{apdx:coneProj}.
Finally, the $\ell_2$ normalization step in line ~\ref{line:normalization} ensures a realistic power profile by scaling the magnitude of the vector without altering its angle and therefore does not compromise the prior projection step. Normalizing the transformed CSI to preserve the norm of the original estimate maintains sample fidelity. Furthermore, although we do not assume a BS equipped with active defense mechanisms, unrealistic power changes could trigger simple anomaly detection algorithms.

It is important to highlight how $T$ and $\alpha$ are directly tied to the cone aperture $\psi_k$, and therefore, to the imposed bound on the SE degradation $\gamma$. For instance, when $\psi_k$ is large, a small gradient step size $\alpha$ will require a larger number of iteration $T$ for the adversarial sample to move outside of the feasible cone. Thus, these hyperparameters cannot be tuned without accounting for the specific system constraints.

\noindent\textbf{\sys within the Reporting Cycle.}
We illustrate the integration of the \sys algorithm within the CSI reporting cycle in Figure~\ref{fig:hideout-scenario}. After estimating the channel $\mathbf{\hat{h}}_k$ and setting a tolerance budget $\gamma$, in \stepnum{2} through \stepnum{4}, the UE computes the CSI report using either the target model or a surrogate, depending on the \textit{white-box} or \textit{black-box} knowledge scenario. The report is computed by ascending the localization loss landscape in directions within the feasible cone $\mathcal{C}_\psi(\mathbf{\hat{h}}_k)$. The BS receives the resulting report in \stepnum{5} and uses it for two parallel tasks: in \stepnum{6a} to locate the UE, and in \stepnum{6b} to beamform the downlink signal. Finally, in \stepnum{7}, the UE receives the downlink service while keeping SE loss within the budget $\gamma$ and successfully obscuring its location from the LS. 

\noindent\textbf{Computer Vision Analogy.}
In adversarial example literature \cite{goodfellow2015explainingharnessingadversarialexamples}, a perturbation is designed to maximize a model's 
loss, while remaining imperceptible to avoid detection. For instance, in image classification, an optimal $\boldsymbol{\delta}^{\text{adv}}$ consistently fools the classifier while the perturbed image remains visually identical to the original by constraining the perturbation's $\ell_\infty$-norm. 

Similarly, in the considered wireless scenario, the main goal remains to maximize the loss. However, unlike in computer vision where constraining the perturbation ensures imperceptibility, our  constraint is designed to maintain a reliable communication link. While a severe drop in link quality or a reported CSI outside of the expected distribution could trigger defense mechanisms, the primary role of the imposed constraint is to guarantee that the UE does not suffer any prohibitive rate costs while obfuscating its location.

\begin{figure}[!t] 
\raggedleft 
\includegraphics[width=\columnwidth]{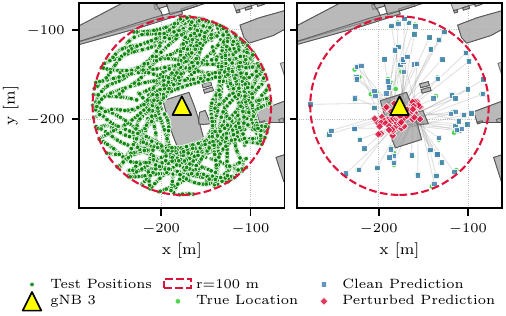}
\caption{Test dataset trajectories and localization accuracy comparison. On the left the spatial distribution of the sampled test positions within a 100\,m radius of gNB~3. On the right an example of \sys mispredicted location in red, with the true locations in blue, clean predictions in green.}
\Description{A Figure with two plots, on the left the sampled locations used to test our methods are shown. On the right, a subset of those sampled locations, with their corresponding clean predictions and the mispredicted locations with \sys under a 5\% maximum drop constraint. The mispredictions can be seen clustering around the gnb.}
\label{fig:combined_data_and_mispreds} 
\end{figure}

\section{Evaluation}
We evaluate whether \sys{} provides location privacy without sacrificing downlink utility. Our evaluation is organized around three questions.

    \noindent\textbf{$\blacktriangleright$} \textbf{RQ1: Privacy--utility tradeoff.} Does \sys{} achieve larger localization error than random or generic adversarial perturbations at the same SE loss?
    
    \noindent\textbf{$\blacktriangleright$}  \textbf{RQ2: Reliability.} Does the protection hold consistently across UE locations rather than only on average?
    
    \noindent\textbf{$\blacktriangleright$} \textbf{RQ3: Transferability.} Do perturbations generated from a local surrogate transfer to unseen target localization architectures in the \textit{black-box} setting?

\subsection{Experimental Setup and Dataset}
To evaluate our approach, we simulate a full link-level protocol in 5G systems, where a UE that wishes to keep its location obscured from the base station computes and injects an adversarial perturbation into the CSI feedback matrix and reports it to the base station uncompressed. 
The base station receives the CSI matrix error-free and, solely relying on it, computes the downlink beamforming vectors. Simultaneously, it also tracks the UE across both line-of-sight (LoS) and non-line-of-sight (NLoS) by feeding the reported channel vectors into its AI-based localization model. 



We evaluate our approach on the open source Multi-cell Outdoor Channel State Information Dataset (MOCSID)~\cite{mocsid}, consisting of channel data generated via NVIDIA Sionna ray-tracing~\cite{sionnaCite}. The dataset models realistic pedestrian trajectories traversing a dense 3D environment of the La Doua campus in Lyon, France. MOCSID utilizes a MISO configuration, where  the UEs are equipped with a single receiving antenna, i.e., $N_{\text{Rx}} = 1$, while the area is covered by 10 base stations equipped with $4 \times 2$ dual-polarized uniform planar arrays, i.e., $N_{\text{Tx}} = 16$. Operating over $K = 256$ OFDM subcarriers, this configuration results in a downlink Channel State Information (CSI) matrix $\mathbf{H} \in \mathbb{C}^{16 \times 256}$. This system-level simulator allows us to draw channel realizations with spatial consistency, i.e., neighboring positions have highly correlated channel matrices. Spatial consistency is guaranteed by the ray-tracing approach of Sionna, which makes it very realistic to measurement data. 

In this work, we focus our analysis on a single-cell scenario, defined as all points within a radius of $r = 100\text{ m}$ from one of the 10 base stations; specifically, we chose the fourth base station. We leave the cross-cell generalizability of these location obfuscation strategies for future work. 

Assuming that the UE is able to estimate the true downlink channel, i.e. $\mathbf{\hat{H}}=\mathbf{H}$, we collect 57,216 downlink channel realizations  $\mathbf{H}$. We then randomly divide the collected data into $85\%$ training, $5\%$ validation, and $10\%$ test sets using a fixed seed. We use the same split to train the target and surrogate localization models, as the same data is available to the UE as well, and report attack results only on the held-out test set.

\subsection{Localization Models}
\label{sec:localization-models}

We deploy four positioning AI-architectures to show that our approach could be model agnostic. All models receive a CSI tensor of shape $N_{\mathrm{Tx}}\times K\times 2$, where the last dimension contains the real and imaginary components. We apply the six-channel transformation described in Section~\ref{sec:ml-positioning} inside each model and use a linear two-neuron output layer to estimate the UE's $(x,y)$ coordinates.

\noindent\textbf{Simple CNN.}
We construct a conventional CNN baseline using seven convolutional layers. The first five layers use $128$ filters, while the last two use $96$ filters. We place max-pooling layers after the second and fourth convolutions, apply global average pooling, and use two fully connected layers with $128$ and $64$ units. Both fully connected layers use a dropout rate of $0.1$.

\noindent\textbf{KUL-CNN.}
We adapt the positioning architecture from~\cite{colpaert2023massive}. The model uses four dense convolutional blocks with kernels of size $1\times9$. The first two blocks use $16$ channels and the final two use $4$ channels. Average-pooling layers reduce the subcarrier dimension between blocks. We then flatten the representation and apply fully connected layers with $128$ and $64$ units, with a dropout rate of $0.1$ after the first layer.

\noindent\textbf{CNN-V3.}
We additionally adapt another CNN model from~\cite{vuckovic2024csi}. It applies five convolutional layers with $8$, $16$, $32$, $32$, and $64$ filters. Their kernel sizes decrease from $8\times8$ to $3\times3$, and batch normalization and max pooling progressively reduce the feature maps before a linear regression layer.

\noindent\textbf{RNN.}
We also include a recurrent model whose structure differs substantially from the three CNNs. We treat the $N_{\mathrm{Tx}}$ antenna vectors as a sequence and flatten the six features over the $K$ subcarriers at each sequence step. Two Long Short-Term Memory (LSTM) layers with $512$ and $64$ units process this sequence. Global average pooling and a $64$-unit fully connected layer produce the representation used for coordinate regression.


We train each target model for at most $100$ epochs using Adam with a learning rate of $10^{-3}$, a batch size of $128$, and mean squared error as the loss. We select the checkpoint with the lowest validation Euclidean error. We halve the learning rate after eight epochs without improvement, down to a minimum of $10^{-6}$, and stop training after $20$ such epochs. These target models serve both as the localization service (LS) models in our \textit{black-box} evaluation and as the directly accessible models in the \textit{white-box} setting.

\subsection{Black-Box Surrogate Search}
\label{sec:surrogate-search}

In the \textit{black-box} setting, the UE does not assume that the LS uses any of the models above. Instead, we use Bayesian optimization to search for a surrogate that 
maps CSI to location. We minimize validation Euclidean positioning error instead of attack performance against a particular target; the search does not use the target model's predictions, parameters, or gradients.

The search covers two broad CNN families. The first consists of standard convolutional blocks, while the second uses densely connected blocks with one-dimensional kernels along the subcarrier axis. Both families use the same six-channel CSI representation as the target models. Table~\ref{tab:surrogate-search-space} summarizes the tunable parameters. Across this search space, the candidate models range from 2.5K to 17M
total parameters. The smallest candidate uses one standard CNN block with one $3\times3$ convolution of $32$ filters and one fully connected layer of $16$ units. The largest candidate uses one dense block with a depth of three, $32$ channels, a $1\times9$ kernel, a pooling width of two, and four fully connected layers of $256$ units each. 

We evaluate $50$ configurations for five epochs each and rank them by validation Euclidean error. We then retrain the three best configurations for at most $100$ epochs, using early stopping with a patience of $10$ epochs, and retain the best validation weights. The resulting model provides the gradients used to generate \textit{black-box} perturbations; the unknown target models only evaluate the perturbed CSI, i.e., $f_{\theta}\left(\mathbf{\hat{H}} + \boldsymbol{\delta}\right)$.

\begin{table}[!t]
    \centering
    \caption{Bayesian search space for the black-box surrogate models.}
    \label{tab:surrogate-search-space}
    \begin{tabular}{ll}
        \toprule
        Parameter & Candidate values \\
        \midrule
        Architecture & standard CNN, dense blocks \\
        Standard CNN blocks & $1$--$5$ \\
        Convolutions per block & $1$--$3$ \\
        Convolution filters & $\{32,64,128\}$ \\
        Convolution kernel & $\{3,5\}$ \\
        Dense blocks & $1$-$4$ \\
        Dense-block depth & $1$-$3$ \\
        Dense-block channels & $\{4,8,16,32\}$ \\
        Dense-block kernel width & $\{3,5,9\}$ \\
        Dense-block pooling width & $\{2,5\}$ \\
        Fully connected layers & $1$--$4$ \\
        Fully connected units & $\{16,64,128,256\}$ \\
        Dropout rate & $\{0.1,0.2,0.3,0.4\}$ \\
        Adam learning rate & $\{10^{-3},5\mathord{\times}10^{-4},10^{-4}\}$ \\
        Total model parameters & $2.5K$--$17M$ \\
        \bottomrule
    \end{tabular}
\end{table}

\begin{figure*}[!t]
    \centering
    
    \begin{minipage}[t]{0.48\textwidth}
        \centering
        \includegraphics[width=\textwidth]{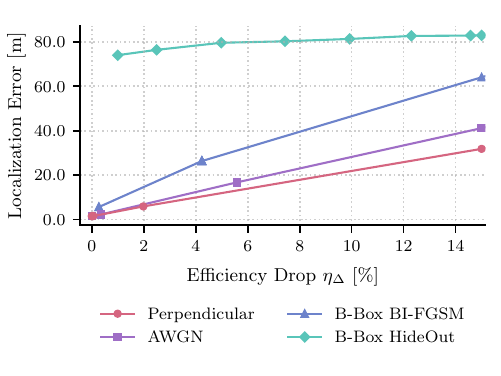}
        \Description{A tradeoff plot showing Average Localization Error in meters on the y-axis (0.0 to 80.0) against Relative SE Drop on the x-axis (0.00 to 0.20). While BI-FGSM forms a curve that slopes down and right, the \sys{} configurations ($\gamma = 1\%$ to $20\%$) cluster tightly in the upper-left corner, showing high error at low efficiency drops.}
        \setsublabel{fig:bbox_error_vs_spectral_efficiency}{a}
        \\ (a) Average localization error vs. SE drop. %
    \end{minipage}
    \hfill
    \begin{minipage}[t]{0.48\textwidth}
        \centering
        \includegraphics[width=\textwidth]{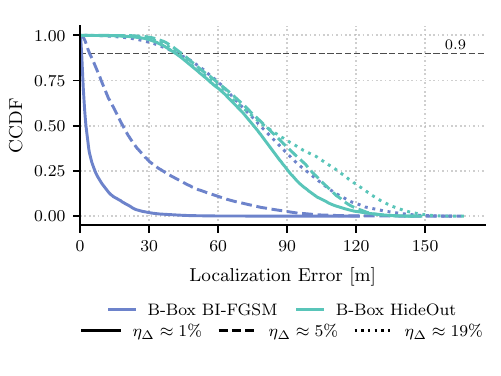}
        \Description{A CCDF plot showing localization error on the x-axis (0 to 150m) and cumulative probability on the y-axis (0 to 1.0). Multiple curves represent \sys{} at varying projection budgets, all of which stay higher and further to the right than the baseline BI-FGSM, indicating higher error rates.}
        \setsublabel{fig:bbox_cdf}{b}
       \\(b) CCDF of localization errors across budgets.%
    \end{minipage}

    \caption{\textit{Black-Box} performance evaluation of \sys{}. (a) We benchmark \sys{} against non-adversarial baselines (AWGN, Perpendicular) and BI-FGSM. \sys{} achieves an optimal tradeoff, maximizing localization error in the upper-left quadrant while strictly bounding downlink degradation. (b) Compared to the BI-FGSM baseline, \sys{} consistently shifts the error distribution to the right. The horizontal dashed line highlights the $0.9$ probability threshold, where \sys{} maintains a localization error of $\ge 40$\,m even under conservative parameters.}
    \label{fig:combined_bbox_KULUEVEN_results}
\end{figure*}

\subsection{Baseline Perturbations}

We benchmark \sys performance against three distinct perturbations.
First, we evaluate \sys against conventional Additive White Gaussian Noise (AWGN). The comparison highlights the advantage of our gradient-based approach over random unstructured isotropic noise.

We then compare \sys against a geometric perturbation, constructed to be strictly orthogonal to the original estimated channel $\mathbf{\hat{H}}$. By inducing the maximum possible geometric rotation to the channel vector, this serves as a worst-case benchmark for downlink disruption. 

Lastly, we compare \sys against the well-known Basic Iterative Fast Gradient Sign Method (BI-FGSM) algorithm~\cite{kurakin2017adversarialexamplesphysicalworld}. At each step $t = 0,\dots,T-1$, the BI-FGSM update is formally defined as:
\begin{equation}
    \label{eq:bifgsm}
    \mathbf{\hat{H}}_{t+1}^{\text{adv}} = \text{Clip}_{\mathbf{\hat{H}}, \epsilon} \Big( 
    \mathbf{\hat{H}}_{t}^{\text{adv}} + \alpha \cdot \mathrm{sign}\big( \nabla_{\mathbf{\hat{H}}} \mathcal{L}(\mathbf{\hat{H}}_{t}^{\text{adv}},\mathbf{r}; \theta) \big) \Big),
\end{equation}
\noindent where $\alpha = \epsilon/T$ and the clipping operator $\text{Clip}_\mathbf{\hat{H}}(\cdot)$ projects the adversarial sample onto an $\ell_\infty$-ball. 

This formulation highlights the parallel to our proposed approach \sys. While BI-FGSM projects the adversary to the $\ell_\infty$-ball at each iteration, \sys projects the perturbation onto the cone $\mathcal{C}_\psi(\mathbf{\hat{h}}_k)$ before normalizing the output norm to match the original estimated channel $\mathbf{\hat{H}}$. Comparing these two methods highlights the benefits of a wireless-context-aware adversarial perturbation. 

\subsection{Performance Metrics}

We assess the impact of our approach using two primary metrics. First, we measure the level of location obfuscation in terms of localization error, i.e. as the Euclidean distance between the ground truth and the predictions of the target model under adversarially perturbed channel realizations. Second, we measure the link degradation due to mismatched beamforming in terms of Shannon SE as defined in Equation~\eqref{eq:effDelta}. 

We then evaluate the reliability of each strategy by considering an obfuscation strategy reliable if it achieves a large localization error consistently across the cell area. We thus compare the Complementary Cumulative Distribution Function (CCDF) of the localization error achieved by the different approaches.

For clarity, we will first focus our detailed tradeoff and CCDF analysis solely on a single model, i.e. KUL-CNN model, operating under a strict \textit{black-box} assumption. We prioritize a \textit{black-box} setting because it represents the most plausible real-world deployment scenario, as base station localization models and weights are typically proprietary. 

Next, we benchmark these \textit{black-box} findings directly against a \textit{white-box} setting, allowing us to assess the impact that having direct target model information has on the efficacy of the perturbation strategy.

Finally, we evaluate how well our strategy generalizes to entirely different models and architectures outside the original surrogate search pool. The complete set of corresponding curves and evaluation figures for the remaining localization models are illustrated in Fig.~\ref{fig:comprehensive_bbox_plots}.

\subsection{Single Model Black-Box Evaluation}



\noindent\textbf{Localization Error and SE Drop Trade-Off.}
Figure~\ref{fig:bbox_error_vs_spectral_efficiency} shows the tradeoff between the Average Localization Error and the SE Drop. While the baseline trajectories are generated by sweeping perturbation budgets $\epsilon$, \sys is evaluated across different constraint configurations $\gamma$.

Both non-adversarial baselines, i.e. AWGN and Perpendicular, yield a poor tradeoff, degrading the SE without providing reliable localization error. This highlights that blindly altering the reported channel disrupts communication performance without guaranteeing a sufficient localization obfuscation level.  

On the other hand, and, as expected, BI-FGSM achieves larger localization errors than non adversarial perturbations. However, as it is not optimized for the considered wireless scenario, BI-FGSM results in high SE degradation.

\sys instead induces a large average localization error even at its most conservative setting. For example, for a strict $\gamma=1\%$, \sys induces an average localization error of $74$\,m. Increasing the efficiency drop budget further increases the average error, which rapidly converges to above $80$\,m. \sys therefore sits within the optimal upper-left quadrant of the tradeoff space, forcing severe tracking errors at a fraction of the efficiency drop compared to baseline perturbations.

\noindent\textbf{CCDF of the Localization Error.}
%
%
Figure~\ref{fig:bbox_cdf} illustrates the CCDF of localization errors with \sys compared to a BI-FGSM baseline configured for comparable efficiency drops under a \textit{black-box} scenario. 

The plotted distributions empirically demonstrate the reliability of \sys, as it enforces large location obfuscation throughout the cell area. Specifically, even for more conservative configurations, e.g. $\gamma \in [1\%, 5\%]$, \sys induces a localization error of $\approx 40$\,m or greater for $90\%$ of the evaluated locations. Increasing the allowable efficiency drop budget shifts the distribution curve to the right, improving the attack's reliability.

Besides highlighting the reliability of \sys, Figure~\ref{fig:bbox_cdf} further highlights its superiority over standard adversarial attacks, as BI-FGSM requires a large efficiency drop budget $\eta_\Delta \approx 19\%$ just to match the performance of \sys operating at a strict $\gamma = 1\%$ constraint.

\begin{figure}[!t]
    \centering
    \includegraphics[width=0.85\columnwidth]{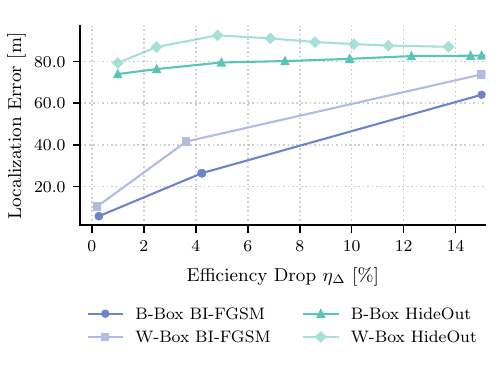}
    \Description{A tradeoff plot showing Average Localization Error in meters on the y-axis (0.0 to 80.0) against Efficiency Drop $\eta_\Delta$ on the x-axis (0\% to 20\%). It plots four lines: B-Box BI-FGSM, W-Box BI-FGSM, B-Box HideOut, and W-Box HideOut. Both knowledge settings for \sys maintain high localization errors between 60m and 80m even at low efficiency drops of 1\% to 5\%, while the BI-FGSM baselines scale up much more slowly.}
        
    \caption{Performance evaluation comparing the obfuscation-efficiency tradeoff under \textit{white-box} (W-Box) and \textit{black-box} (B-Box) knowledge settings. \textit{Black-Box} tracking closely mirrors the \textit{white-box} trends; however, the \textit{white-box} setting achieves a strictly larger localization error at equivalent efficiency drops by leveraging exact target model gradients. }
    \label{fig:wbox_tradeoff}
\end{figure}

\noindent\textbf{\textit{White-Box} Versus \textit{Black-Box}.}
Figure~\ref{fig:wbox_tradeoff} compares the two different knowledge settings by plotting their obfuscation-efficiency tradeoff. Both gradient-based perturbations, i.e. BI-FGSM and \sys, maintain their performance trends across both knowledge settings. However, both achieve a better tradeoff in a \textit{white-box} setting, as the perturbation is computed with full model access.

\subsection{\sys Across Different Models}
\label{sec:generalizability}

\begin{figure*}[!t]
    \centering
    
    \begin{minipage}[t]{0.32\textwidth}
        \centering
        \includegraphics[width=\textwidth]{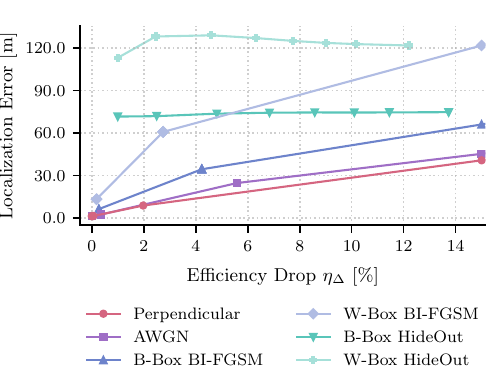}
        \Description{A tradeoff plot for Simple CNN showing the average localization error in meters on the y-axis against relative SE drop on the x-axis ranging from 0 to 20 percent. The \sys curve remains high and flat above 62 meters across all efficiency drops, vastly outperforming the lower, gradually rising baseline curves. The baseline BI-FGSM only reaches and surpasse \sys for an efficiency drop larger than 17\%}
        \setsublabel{fig:simple_cnn_bbox_tradeoff}{a}
        \\ \small (a) Simple CNN \textit{black-box} Trade-off
    \end{minipage}
    \hfill
    \begin{minipage}[t]{0.32\textwidth}
        \centering
        \includegraphics[width=\textwidth]{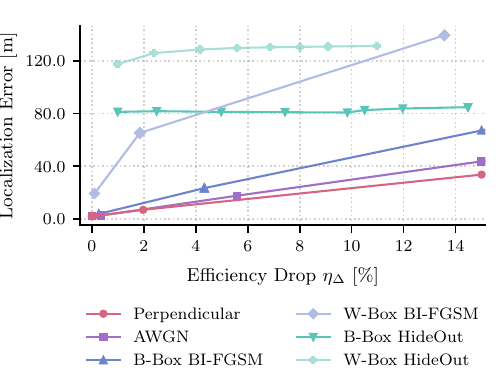}
        \Description{A tradeoff plot for CNN-V3 showing the average localization error in meters on the y-axis against relative SE drop on the x-axis ranging from 0 to 20 percent. The \sys curve sits flat and dominant in the top left corner of the plot, demonstrating strong performance advantages over non-adversarial baselines at low overhead levels.The baseline BI-FGSM only reaches and surpasses \sys for an efficiency drop larger than 17\%}
        \setsublabel{fig:twc24_cnn_bbox_tradeoff}{b}
        \\ \small (b) CNN-V3 \textit{black-box} Trade-off
    \end{minipage}
    \hfill
    \begin{minipage}[t]{0.32\textwidth}
        \centering
        \includegraphics[width=\textwidth]{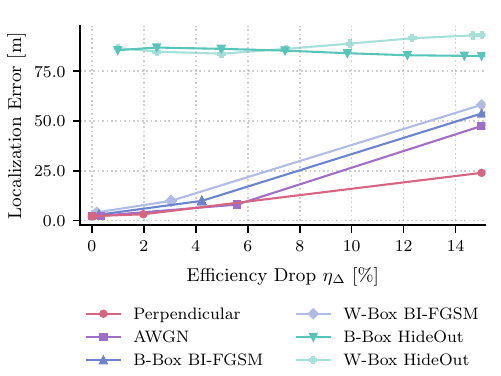}
        \Description{A tradeoff plot for the RNN architecture  showing the average localization error in meters on the y-axis against relative SE drop on the x-axis ranging from 0 to 20 percent. The \sys curve runs high and flat near the 78-meter mark across the entire horizontal axis, completely eclipsing the BI-FGSM baseline curve which never surpasses 60m of error.}
        \setsublabel{fig:rnn_bbox_tradeoff}{c}
        \\ \small (c) RNN \textit{black-box} Trade-off
    \end{minipage}

    \begin{minipage}[t]{0.32\textwidth}
        \centering
        \includegraphics[width=\textwidth]{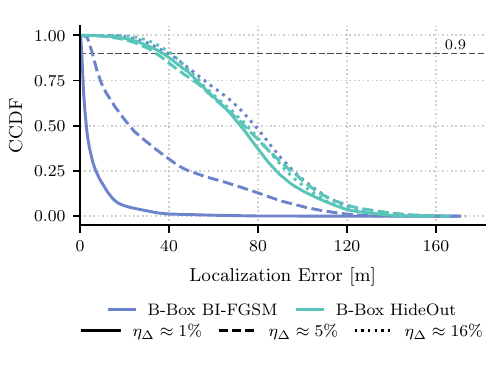}
        \Description{A CCDF plot for Simple CNN showing localization error in meters on the x-axis from 0 to 150m and cumulative probability on the y-axis from 0.00 to 1.00. The curves trace the spatial distribution of errors induced by \sys across various configuration budgets from 1.00 percent to 18.68 percent against BI-FGSM baselines with comparable drops. BI-FGSM results in a similar CCDF only for efficiency drop above 15\%}
        \setsublabel{fig:simple_cnn_bbox_ccdf}{d}
        \\ \small (d) Simple CNN \textit{black-box} CCDF
    \end{minipage}
    \hfill
    \begin{minipage}[t]{0.32\textwidth}
        \centering
        \includegraphics[width=\textwidth]{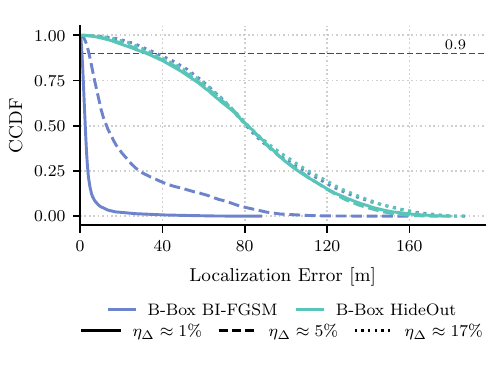}
        \Description{A CCDF plot for CNN-V3 showing localization error from 0 to 150 meters on the x-axis and cumulative probability from 0.00 to 1.00 on the y-axis. The curves representing all \sys configuration budgets cluster tightly together. BI-FGSM achieves a comparable CCDF only for larger efficiency drops above 15\%}
        \setsublabel{fig:twc24_cnn_bbox_ccdf}{e}
        \\ \small (e) CNN-V3 \textit{black-box} CCDF
    \end{minipage}
    \hfill
    \begin{minipage}[t]{0.32\textwidth}
        \centering
        \includegraphics[width=\textwidth]{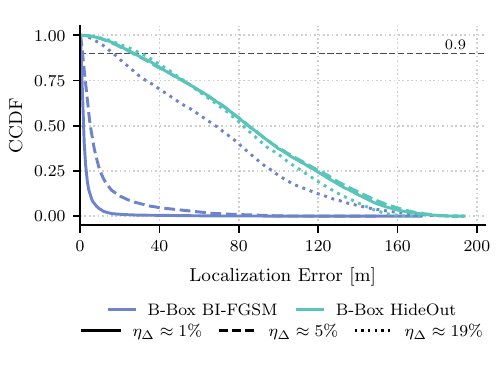}
        \Description{A CCDF plot for the RNN architecture showing localization error from 0 to over 150 meters on the x-axis and cumulative probability from 0.00 to 1.00 on the y-axis. The tightly clustered curves for the \sys configurations are positioned significantly to the right of the orange BI-FGSM baseline curve, indicating higher overall tracking errors across the cell.}
        \setsublabel{fig:rnn_bbox_ccdf}{f}
        \\ \small (f) RNN \textit{black-box} CCDF
    \end{minipage}

    \caption{Comprehensive empirical results summary across remaining localization architectures. Subplots (a–c) depict the obfuscation-efficiency tradeoff, where the \sys curve remains characteristically high and flat across all overhead levels (e.g., maintaining errors above $70$\,m across architectures), significantly outperforming the baselines. For instance, the BI-FGSM baseline only approaches \sys at efficiency drops larger than $14\%$. The plots also highlight the clear drawback of the \textit{black-box} setting, however the trends remain preserved across knowledge setting. Subplots (d–f) show the CCDF of localization errors across $1.00\% \le \eta_\Delta \le 20\%$, illustrating that the tightly clustered \sys configurations consistently shift the error distribution to the right compared to the baseline, which requires substantial efficiency drops larger than 16\% to achieve comparable degradation in the CNN variants.}
    
    \label{fig:comprehensive_bbox_plots}
\end{figure*}

We evaluate the transferability of \sys across the four localization architectures described in Section~\ref{sec:localization-models}. 
Figure~\ref{fig:comprehensive_bbox_plots} gives a comprehensive overview of the corresponding tradeoff and CCDF curves. These empirical results show a persistent cross-architecture transferability of both optimization-based perturbations, i.e. standard BI-FGSM and \sys.

\noindent\textbf{Localization Error and SE Drop Trade-off.}
As shown in Figures~\ref{fig:simple_cnn_bbox_tradeoff}, ~\ref{fig:twc24_cnn_bbox_tradeoff} and ~\ref{fig:rnn_bbox_tradeoff}, our proposed approach dominates the baselines along the obfuscation-efficiency tradeoff across all various model architectures, 
even under  stringent efficiency constraints. For instance, in a \textit{black-box} setting \sys forces an average tracking error greater than $70$\,m across all of the explored architectures, even with $\gamma\leq5\%$. In contrast, standard AWGN and BI-FGSM  can only achieve comparable levels of localization obfuscation for efficiency drop budgets larger than 14\%. 

The high cross-structural generalizability of \sys is further highlighted in Figure~\ref{fig:rnn_bbox_tradeoff}, which displays the tradeoff achieved on an RNN architecture. On the evaluated RNN, \sys induces an average error of $84.96$\,m under a strict $\gamma=1\%$, completely dominating the $2.91$\,m achieved by BI-FGSM. The performance of \sys remains stable across the entire continuous spectrum of efficiency drops. On the other hand, BI-FGSM never exceeds $60$\,m of localization error, even as the efficiency budget relaxes. These results highlight the higher transferability of \sys, as the RNN belongs to an entirely different model family than the CNNs utilized within the surrogate search pool.

\noindent\textbf{CCDF of the Localization Error.}
Across all three evaluation architectures, the error distributions for the various \sys configurations cluster tightly together, as shown in Figures~\ref{fig:simple_cnn_bbox_ccdf}, ~\ref{fig:twc24_cnn_bbox_ccdf}, and ~\ref{fig:rnn_bbox_ccdf}. In contrast, the BI-FGSM baseline requires an efficiency drop greater than $16\%$ to achieve a comparable rightward shift in its error distribution. The tight distribution of \sys demonstrates once more how our proposed approach delivers consistent, wide-area localization degradation, compared to the baselines.

\noindent\textbf{\textit{White-Box} Versus \textit{Black-Box}.}
Figures~\ref{fig:simple_cnn_bbox_tradeoff} and ~\ref{fig:twc24_cnn_bbox_tradeoff} highlight once more the advantage of having full model access, as both gradient-based perturbations achieve a better tradeoff in a \textit{white-box} setting. However, as displayed in Figure~\ref{fig:rnn_bbox_tradeoff}, while having precise gradient information provides a clear edge when attacking the CNN target, the advantage shrinks significantly when evaluated against the RNN target. This seems to be caused by rather noisy gradients in deep RNNs compared to other architectures.

\section{Limitations and Deployment Considerations}
\label{sec:discussion}
Our evaluation considers localization models trained on clean CSI and tested on protected CSI. A network-side adversary could attempt to harden the localizer through adversarial training, detect distribution shifts in CSI reports, or fuse CSI with independent positioning or classical sources. \sys{} should therefore be viewed as a client-side protection mechanism against CSI-only localization, not as a complete defense against all forms of cellular tracking. Extending the defense to adaptive and multi-modal localization services is an important direction for future work.

Computing \sys in the real world requires the UE to locally store and execute a trained surrogate localization model. Offline training is not a major bottleneck, as necessary training data can be collected prior to deployment or synthesized using channel simulators, and then the model can be fine-tuned using on-device measurements. 

Non-volatile memory cost is also not a major limiting factor when executing \sys in real-world deployments. For instance, in the surrogate pool we used for evaluation, models ranged from 2,418 to 17,069,538 parameters. Under a standard 32-bit single-precision floating-point (\texttt{fp32}) representation, this corresponds to a local memory footprint ranging from approximately $9.45\text{ KiB}$ to $65.12\text{ MiB}$, which modern commercial off-the-shelf (COTS) smartphones can easily accommodate.

Real-time execution under network timing constraints, however, poses the primary challenge for real-world deployments. Computing \sys, in fact, requires the execution of an initial forward pass followed by alternating gradient updates and cone projections. 
These operations must be executed within the typical reporting intervals of the CSI feedback loop for each new location. However, CSI reporting, which can be configured as periodic, semi-persistent, or aperiodic \cite{3gpp38214}, uses configurable reporting periodicities ranging from 4 to 320 slots, corresponding to time intervals from approximately 0.5 ms to 320 ms depending on the configured numerology.
As \sys execution time scales with the dimensions of $\mathbf{\hat{H}} \in \mathbb{R}^{N_{\text{Tx}} \times K \times 2}$, the surrogate model parameter count, and the number of iterations $T$, for highly mobile scenarios or with deep architectures, \sys can introduce a non-negligible hardware strain. 

\begin{table}[t]
\centering
\caption{Runtime comparison between BI-FGSM and the fastest HideOut implementation. Overhead is computed relative to BI-FGSM.}
\label{tab:runtime_comparison}
\small
\setlength{\tabcolsep}{1.5pt}
\begin{tabular}{@{}ccrrrr@{}}
\toprule
\rotatebox[origin=c]{-45}{\textbf{Model}} &
\rotatebox[origin=c]{-45}{\textbf{Params}} &
\rotatebox[origin=c]{-45}{\textbf{Iter.}} &
\rotatebox[origin=c]{-45}{\textbf{BI-FGSM (ms)}} &
\rotatebox[origin=c]{-45}{\textbf{Fastest HideOut (ms)}} &
\rotatebox[origin=c]{-45}{\textbf{Overhead (\%)}} \\
\midrule
\multirow{4}{*}{\rotatebox[origin=c]{-45}{KUL-CNN}}
& \multirow{4}{*}{69 K}
& 1  & 6.1   & 10.9  & 78.7 \\
& & 10 & 58.2  & 103.4 & 77.7 \\
& & 20 & 113.3 & 196.2 & 73.1 \\
& & 30 & 169.4 & 294.9 & 74.1 \\
\midrule
\multirow{4}{*}{\rotatebox[origin=c]{-45}{CNN-V3}}
& \multirow{4}{*}{83 K}
& 1  & 9.4   & 14.4  & 52.9 \\
& & 10 & 89.2  & 125.8 & 41.0 \\
& & 20 & 170.5 & 254.9 & 49.5 \\
& & 30 & 254.6 & 411.7 & 61.7 \\
\midrule
\multirow{4}{*}{\rotatebox[origin=c]{-45}{Simple CNN}}
& \multirow{4}{*}{741 K}
& 1  & 52.2  & 56.7  & 8.7 \\
& & 10 & 513.3 & 555.5 & 8.2 \\
& & 20 & 1024.9 & 1111.1 & 8.4 \\
& & 30 & 1538.5 & 1665.2 & 8.2 \\
\midrule
\multirow{4}{*}{\rotatebox[origin=c]{-45}{RNN}}
& \multirow{4}{*}{4.35 M}
& 1  & 22.4  & 27.2  & 21.7 \\
& & 10 & 217.0 & 256.8 & 18.3 \\
& & 20 & 434.3 & 508.3 & 17.0 \\
& & 30 & 686.6 & 779.5 & 13.5 \\
\midrule
\multirow{4}{*}{\rotatebox[origin=c]{-45}{Surrogate}}
& \multirow{4}{*}{1.01 M}
& 1  & 16.9  & 21.7  & 28.7 \\
& & 10 & 166.4 & 211.1 & 26.9 \\
& & 20 & 333.5 & 414.4 & 24.3 \\
& & 30 & 498.2 & 615.6 & 23.6 \\

\bottomrule
\end{tabular}
\end{table}

To quantify this processing overhead on resource-constrained hardware, we evaluated \sys on a virtual machine configured with 2\,GiB of RAM and 2 vCPU cores running Ubuntu 24.04 LTS. To faithfully simulate low-power edge execution on a COTS UE, execution was strictly constrained to a single-threaded, CPU-only environment. Table~\ref{tab:runtime_comparison} summarizes the single-thread execution latency across varying iteration counts ($T \in \{1,10, 20, 30\}$), comparing BI-FGSM against the fastest \sys implementation. While lightweight surrogates with lower iteration counts can fit within moderate CSI reporting cycles, under high UE mobility, scaling to deeper architectures clearly imposes a bottleneck relative to network timing constraints.

Addressing these constraints remains an active research direction; we are currently developing a lightweight, resource-optimized variant of \sys for future deployment scenarios.



%

%
\section{Related Work}

\noindent\textbf{ML-based localization from CSI.}
Recent work has shown that CSI contains sufficient
location-dependent information for accurate positioning.
CNN-based systems can directly estimate UE coordinates
from massive-MIMO CSI measurements
~\cite{colpaert2023massive,vuckovic2024csi},
while integrated feedback-positioning architectures
jointly optimize CSI reconstruction and localization
~\cite{guo2024integrated}.
These systems demonstrate that location information is
embedded in CSI and can be extracted by machine-learning
models. We instead ask how a UE can suppress such
information before reporting CSI, while preserving the
report's utility for downlink precoding.

\noindent\textbf{Location privacy above the physical layer.}
Most location-privacy mechanisms protect coordinates after localization has already occurred. 
Geo-indistinguishability adds calibrated noise to reported locations so that nearby positions are difficult to distinguish~\cite{andres2013geo}. 
L-SRR uses local differential privacy for location-based services~\cite{wang2022srr}. 
At the cellular-network level, Pretty Good Phone Privacy reduces location tracking by decoupling long-term subscriber identifiers from network access~\cite{schmitt2021pgpp}. 
These mechanisms protect released locations or identifiers, but they do not prevent a network-side localization service from inferring a UE's position directly from raw CSI used by the communication system.

\noindent\textbf{Signal-level privacy in wireless systems.}
A smaller body of work acts directly on wireless signals or measurements. 
For geometry-based cellular localization, Checa and Tomasin design a privacy-preserving beamformer that hides angle-of-departure information at the cost of achievable rate~\cite{Checa2020LocationPrivacy}, while Khan~\emph{et al.} reshape angular power so that a false direction appears dominant~\cite{Khan2025DoAOPR}. 
These works expose the coupling between localization and communication when both depend on explicit geometric channel features.

Most signal-level privacy work, however, targets WiFi or short-range sensing. 
Cominelli~\emph{et al.} randomize transmitted WiFi I/Q samples to disrupt CSI-based device-free localization while preserving communication~\cite{cominelli2020csi}. 
Zhou~\emph{et al.} learn adversarial transformations that remove selected behavioral information from WiFi CSI while retaining information needed by another classifier~\cite{zhou2020adversarialwifi}. 
FooLoc modifies WiFi training symbols so that an access point obtains misleading CSI fingerprints~\cite{xiao2023fooloc}, and Li~\emph{et al.} perturb WiFi packet preambles to manipulate learning-based sensing without interrupting communication~\cite{li2024practicalwifi}. 
These systems demonstrate that wireless signals can be manipulated for privacy or attack, but they do not address the cellular feedback loop in which the same CSI report is consumed by both a localization service and a downlink precoder.

\noindent\textbf{Adversarial perturbations and black-box transfer.}
\sys is related to adversarial machine learning, where perturbations are optimized to increase a model's loss while satisfying a constraint on the modified input~\cite{kurakin2017adversarialexamplesphysicalworld}. 
\textit{Black-box} attacks often rely on surrogate models and transferability of adversarial examples~\cite{papernot2017practical,liu2017delving}. 
However, standard adversarial attacks constrain perturbations with generic norms such as $\ell_\infty$, which do not capture the physical effect of CSI perturbations on beamforming. 
\sys differs by replacing the generic norm constraint with a wireless feasibility constraint derived from the downlink SE budget.

\noindent\textbf{Positioning of \sys.}
Prior work establishes that CSI can support accurate localization and that wireless measurements can be transformed to reduce sensing leakage. 
However, we did not find prior work that lets a UE protect its own location by perturbing its cellular CSI feedback while enforcing a selected downlink-rate budget. 
\sys addresses this setting directly: the UE protects its location before CSI disclosure, the localization service processes the protected report, and the base station still uses the same report for downlink precoding.

\section{Conclusions}
\label{sec:conclusion}

We presented \sys{}, a client-side defense against CSI-based localization in mobile networks. \sys{} perturbs CSI feedback to increase the error of neural localization models while enforcing a downlink SE budget derived from the wireless link. By converting the rate-loss constraint into a feasibility set under MRT precoding, \sys{} replaces generic adversarial norm constraints with a communication-aware projection step.

Our evaluation shows that \sys{} substantially degrades CSI-based localization across multiple neural architectures and under \textit{black-box} transfer while keeping rate loss within user-selected budgets. These results suggest that mandatory CSI feedback can be protected before disclosure, but they also highlight a broader lesson: privacy defenses for AI-enabled wireless systems must account for the physical layer utility of the data they perturb.

\begin{acks}
This work has been partially supported by the BMFTR project 6G-life (16KIS2414). 
The work of Jafar Aco Mohammadi is co-funded by the European Union (EU) under Grant Agreement
101191936. Views and opinions expressed are, however, those of the authors only and do not necessarily reflect those of all SUSTAIN-6G consortium parties nor those of the EU or the SNS JU. Neither the EU nor the granting authority (SNS JU) can be held responsible for them.
\end{acks}

\bibliographystyle{ACM-Reference-Format}
\bibliography{references}

\appendix


\section{Cone Analysis}
\subsection{Relation Between \texorpdfstring{$\Gamma_k$}{Gamma} and \texorpdfstring{$\psi_k$}{Psi}}
\label{apdx:gammaPsi}
For the following derivations we assume unit transmit power, i.e., $p=1$. Additionally, we denote by $\rho_k :=\|\mathbf{h}_k\|^2/\sigma^2$ the SNR of the true channel, by $\beta_k$ the Hermitian angle between the true channel and its estimate, and by $\psi_k$ the Hermitian angle between the true channel and the reported one. The SE degradation induced by a mismatched precoder can be therefore written as
\begin{align*}
    \Gamma_k &= \eta\left(\hat{\mathbf{h}}_k\right) - \eta\left(\hat{\mathbf{h}}_k + \boldsymbol{\delta}_k\right) \\
    &= \log_2 \left( 1 + \frac{|\langle \mathbf{h}_k, \mathbf{w}(\hat{\mathbf{h}}_k) \rangle|^2}{\sigma^2} \right) - \log_2 \left( 1 + \frac{|\langle \mathbf{h}_k, \mathbf{w}(\hat{\mathbf{h}}_k + \boldsymbol{\delta}_k) \rangle|^2}{\sigma^2} \right).
\end{align*}
Under MRT precoding, the above relation can be written as  
\begin{align*}
    &2^{\Gamma_k}\left( 1 + \rho_k \cos^2 \psi_k \right)=1 + \rho_k\cos^2 \beta_k , \\
    &\cos^2\psi_k = \frac{\cos^2 \beta_k - (2^{\Gamma_k}-1)/\rho_k}{2^{\Gamma_k}}.
\end{align*}
Given that $\Gamma_k = \gamma \, \eta(\hat{\mathbf{h}}_k)$ with $\gamma \in (0,1)$, then $2^{\Gamma_k} = \left(1 + \rho_k\cos^2\beta_k\right)^\gamma$. Thus,
\begin{equation*}
    \cos^2\psi_k =\frac{\left(1+\rho_k \cos^2\beta_k\right)^{1-\gamma} -1}{\rho_k},
\end{equation*}
and therefore,
\begin{equation*}
        \psi_k = \arccos \left( \sqrt{\frac{\left(1+\rho_k \cos^2\beta_k\right)^{1-\gamma} -1}{\rho_k}} \right).
\end{equation*}
Since $\gamma \in (0,1)$ and $1+\rho_k\cos^2\beta_k \geq 1$, the radicand is non-negative and $\psi_k$ is always well defined.

\begin{figure*}[!t] 
\centering 
\begin{subfigure}[t]{0.32\textwidth} 
\centering 
\includegraphics[width=\linewidth]{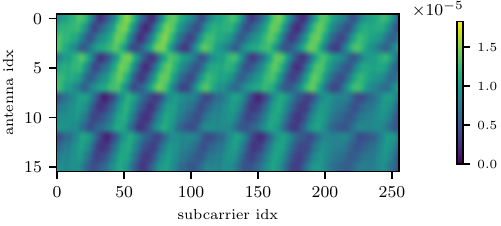}
\Description{Heatmap of a clean CSI matrix with distinct diagonal stripes.} 
\caption{Clean CSI} \label{fig:clean_csi} \end{subfigure} 
\hfill 
\begin{subfigure}[t]{0.32\textwidth} 
\centering 
\includegraphics[width=\linewidth]{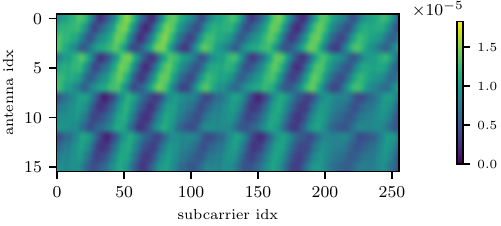}
\Description{Heatmap of a CSI matrix perturbed with \sys, retaining the diagonal features of the clean CSI matrix.} 
\caption{\sys-perturbed CSI} 
\label{fig:hideout_csi} 
\end{subfigure} 
\hfill 
\begin{subfigure}[t]{0.32\textwidth} 
\centering 
\includegraphics[width=\linewidth]{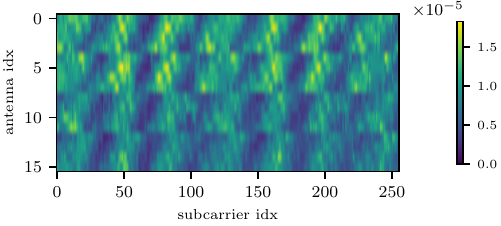}
\Description{Heatmap of a CSI matrix perturbed with BI-FGSM, showing high-intensity unstructured noise.} 
\caption{BI-FGSM-perturbed CSI} \label{fig:bifgsm_csi} 
\end{subfigure} 
\caption{Visual comparison of the absolute CSI matrices under different perturbation conditions. (a) The clean CSI matrix exhibits distinct diagonal structures. (b) The CSI matrix perturbed with \sys maintains high visual fidelity to the clean sample and preserves its characteristic structures. (c) Under the same perturbation budget, BI-FGSM introduces unstructured noise that alters the original CSI characteristics.} 
\label{fig:csi_visual_comparison} 
\end{figure*}






\subsection{Projection onto a Complex Cone}
\label{apdx:coneProj}

Consider the set
\begin{equation*}
\mathcal{C}_\psi(\mathbf{h}) =
\left\{
\mathbf{x} \in \mathbb{C}^{N_{\text{Tx}}}
\;\middle|\;
\frac{|\mathbf{h}^H \mathbf{x}|}{\|\mathbf{h}\| \|\mathbf{x}\|} \ge \cos\psi_k
\right\}.
\end{equation*}
Given $\mathbf{w} \in \mathbb{C}^{N_{\text{Tx}}}$, we define
\begin{equation*}
\mathcal{P}_{\mathcal{C}_\psi}(\mathbf{w}) :=
\arg\min_{\mathbf{x}\in\mathcal{C}_\psi(\mathbf{h})}
\|\mathbf{x}-\mathbf{w}\|^2,
\end{equation*}
as the projection of $\mathbf{w}$ onto $\mathcal{C}_\psi(\mathbf{h})$.
We distinguish two cases:
\begin{enumerate}
    \item Interior case
    If $\mathbf{w} \in \mathcal{C}_\psi(\mathbf{h})$, then
    \begin{equation*}
    \mathcal{P}_{\mathcal{C}_\psi}(\mathbf{w}) = \mathbf{w}.
    \end{equation*}

    \item Boundary case
    If $\mathbf{w} \notin \mathcal{C}_\psi(\mathbf{h})$, then the optimal solution satisfies
    \begin{equation*}
    \frac{|\mathbf{h}^H \mathcal{P}_{\mathcal{C}_\psi}(\mathbf{w})|}{\|\mathbf{h}\| \|\mathcal{P}_{\mathcal{C}_\psi}(\mathbf{w})\|} = \cos\psi_k,
    \end{equation*}
    i.e. lies on the boundary of the cone and, 
    \begin{equation*}
    \mathcal{P}_{\mathcal{C}_\psi}(\mathbf{w}) \in \mathrm{span}\{\mathbf{h},\mathbf{w}\},
    \end{equation*}
    i.e. lies on the plane defined by $\mathbf{w}$ and $\mathbf{h}$
    \end{enumerate}
We now construct an orthonormal basis for $\mathrm{span}\{\mathbf{h},\mathbf{w}\}$:
\begin{equation*}
\mathbf{e}_1 = \frac{\mathbf{h}}{\|\mathbf{h}\|},
\qquad
\mathbf{e}_2 =
\frac{\mathbf{w} - (\mathbf{e}_1^H \mathbf{w})\mathbf{e}_1}
{\left\|\mathbf{w} - (\mathbf{e}_1^H \mathbf{w})\mathbf{e}_1\right\|}.
\end{equation*}
Since $\mathbf w\notin\mathcal C_\psi(\mathbf h)$,
$\mathbf w$ cannot be collinear with $\mathbf h$ and
$\mathbf e_2$ is therefore well defined.
Any vector in this subspace can be written as
$
\mathbf{x} = a \mathbf{e}_1 + b \mathbf{e}_2,
\; a,b \in \mathbb{C},
$ where 
$
a = \mathbf{e}_1^H \mathbf{x}$ and $b = \mathbf{e}_2^H \mathbf{x}.
$
Substituting into the constraint yields
\begin{equation*}
\frac{|a|}{\sqrt{|a|^2 + |b|^2}} = \cos\psi_k,
\end{equation*}
which implies the boundary relation $|b| = |a| \,\tan\psi_k.$
%
%
Hence, all boundary feasible directions can be parameterized as
\begin{equation*}
\mathbf{u}(\phi)
=
\cos\psi_k \mathbf{e}_1
+
\sin\psi_k\, e^{j\phi} \mathbf{e}_2,
\qquad \phi \in [0,2\pi).
\end{equation*}
The projection is obtained by selecting the boundary direction that minimizes the distance to $\mathbf{w}$:
\begin{equation*}
\phi^\star =
\arg\max_{\phi \in [0,2\pi)}
\left| \mathbf{u}(\phi)^H \mathbf{w} \right|.
\end{equation*}
For the optimal direction $\mathbf{u}(\phi^\star)$, the optimal scaling is
\begin{equation*}
r^\star = \left| \mathbf{u}(\phi^\star)^H \mathbf{w} \right|.
\end{equation*}
The projection is therefore given by
$
\mathcal{P}_{\mathcal{C}_\psi}(\mathbf{w}) = r^\star \mathbf{u}(\phi^\star).$

\section{Additional Results}

This appendix provides supplementary results to further validate and illustrate the characteristics of our localization framework. 

\subsection{CSI Feature Visualization under Attack}
To gain a deeper qualitative understanding of how \sys perturbs the CSI compared to baselines, we visualize the reported CSI as heatmaps, where the intensity is proportional to the absolute value of each matrix entry. The visualization gives insight into how \sys affects the reported CSI against a standard adversarial perturbation, i.e., BI-FGSM.

\subsection{Further Cross-Model Generalization Evaluation}
Table~\ref{tab:comprehensive_Results_1} evaluates \sys{}'s \textit{black-box} transferability under 
$\gamma=1\%$ efficiency constraint across the four localization architectures and compares it to  the other baselines. 
$P_{90}$ refers to the localization error that is met or
exceeded at $90\%$ of the evaluated locations, i.e. the point at which the CCDF crosses $0.9$. The table highlights the clear superiority of our proposed approach under this strict budget. For example, on the Simple CNN, \sys forces a massive average tracking error of $71.6$\,m, whereas standard AWGN and BI-FGSM degrade accuracy by at most $2.25$\,m and $6.10$\,m, respectively. 

\begin{table}[!hbt]
\centering
\caption{Tradeoff comparison of perturbations against different architectures for a service drop $\eta_\Delta=1\%$}
\label{tab:comprehensive_Results_1}
\resizebox{\columnwidth}{!}{%
\begin{tabular}{c ccc c}
\toprule
\multirow{2}{*}{\shortstack[c]{\textbf{Model} \\ (Clean Avg.)}} & \multirow{2}{*}{\textbf{Metric}} & \multicolumn{3}{c}{\textbf{$\eta_\Delta=1\%$}} \\
\cmidrule(lr){3-5}
& & \textbf{AWGN} & \textbf{BI-FGSM} & \textbf{\sys} \\
\midrule 
\multirow{3}{*}{\shortstack[c]{\textbf{KUL-CNN} \\ (1.54 m)}} 
& Avg. Err. [m]      &   2.33 &   5.66 &  74.02 \\
& $P_{90}$ Err. [m]  &   0.43 &   0.68 &  41.35 \\
& $\eta_\Delta$ [\%] &   0.35 &   0.27 &   1.00 \\
\midrule
\multirow{3}{*}{\shortstack[c]{\textbf{CNN-V3} \\ (2.11 m)}} 
& Avg. Err. [m]      &   2.61 &   3.82 &  81.33 \\
& $P_{90}$ Err. [m]  &   0.73 &   0.90 &  32.08 \\
& $\eta_\Delta$ [\%] &   0.35 &   0.27 &   1.00 \\
\midrule
\multirow{3}{*}{\shortstack[c]{\textbf{Simple CNN} \\ (1.08 m)}} 
& Avg. Err. [m]      &   2.25 &   6.10 &  71.60 \\
& $P_{90}$ Err. [m]  &   0.34 &   0.75 &  37.09 \\
& $\eta_\Delta$ [\%] &   0.35 &   0.27 &   1.00 \\
\midrule
\multirow{3}{*}{\shortstack[c]{\textbf{RNN} \\ (2.15 m)}} 
& Avg. Err. [m]      &   2.36 &   2.91 &  84.96 \\
& $P_{90}$ Err. [m]  &   0.49 &   0.52 &  25.61 \\
& $\eta_\Delta$ [\%] &   0.35 &   0.27 &   1.00 \\
\bottomrule
\end{tabular}%
}
\end{table} 



\section*{AI Use Declaration}

Generative AI tools were not used to conceive the main research idea underlying this paper, develop the proposed methodology, design the core algorithms, determine the experimental design, or derive the scientific conclusions. These aspects were conceived and developed by the authors.
However, LLM-based tools have been used in two limited ways during this work. First, AI-coding tools were used to help implement auxiliary functions, refactor code, and optimize runtime performance. Second, generative AI was partially used as a writing assistant to suggest improvements to grammar, clarity, organization, and wording. All technical content, scientific claims, and the final manuscript text were reviewed, revised, and verified by the authors, who take full responsibility for the content of this paper.

\end{document}